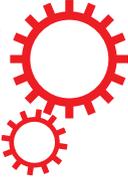

# Geometric frustration in ordered lattices of plasmonic nanoelements

Ana Conde-Rubio[1,2], Arantxa Fraile Rodríguez[1,2], André Espinha[3], Agustín Mihi[3], Francesc Pérez-Murano[4], Xavier Batlle[1,2] & Amílcar Labarta[1,2]

Inspired by geometrically frustrated magnetic systems, we present the optical response of three cases of hexagonal lattices of plasmonic nanoelements. All of them were designed using a metal-insulator-metal configuration to enhance absorption of light, with elements in close proximity to exploit near-field coupling, and with triangular symmetry to induce frustration of the dipolar polarization in the gaps between neighboring structures. Both simulations and experimental results demonstrate that these systems behave as perfect absorbers in the visible and/or the near infrared. Besides, the numerical study of the time evolution shows that they exhibit a relatively extended time response over which the system fluctuates between localized and collective modes. It is of particular interest the echoed excitation of surface lattice resonance modes, which are still present at long times because of the geometric frustration inherent to the triangular lattice. It is worth noting that the excitation of collective modes is also enhanced in other types of arrays where dipolar excitations of the nanoelements are hampered by the symmetry of the array. However, we would like to emphasize that the enhancement in triangular arrays can be significantly larger because of the inherent geometric incompatibility of dipolar excitations and three-fold symmetry axes.

Frustration has largely been studied in magnetism where the term refers to situations in which one or more spins in the magnetic unit cell do not find a proper orientation to fully satisfy all the interactions with the neighboring spins, either by the existence of competing interactions or by the actual geometry of the structure itself[1,2]. For example, this phenomenon typically occurs in systems with hexagonal-based lattices and short range antiferromagnetic interactions, where geometric frustration yields slow dynamics and ageing effects both caused by an intricate energy landscape with lots of quasi-degenerate states[3,4].

This article aims at exploiting the concept of geometric frustration in the field of plasmonics. We introduced this term in a recent paper, referring to situations in which the geometry of a plasmonic lattice does not favor a complete ferroelectric polarization of the gaps between neighboring elements[5]. In particular, we showed the case of a hexagonal lattice of asterisks that contained three sub-lattices of gaps with different orientations with respect to the incident field, in such a way that the interactions among the gap dipoles destabilize the mode corresponding to the full polarization of all the gaps.

Here, we take a leap forward and address the issue of the actual underlying mechanism behind the effect. In particular, we focus on the study of other arrays of Au nanoelements in which near-field dipolar polarizations of neighboring elements are geometrically frustrated. Our goal is to favor the excitation of collective modes at the expense of the low energy modes corresponding to near-field interactions with prevailing dipolar character. We show that the key ingredients in settling the degree of geometric frustration of these lattices depend on the complexity of the system, that is, the number of sublattices of gaps with different spatial orientations, together with the strength of the interactions among them. The former is related to the symmetry properties of both the lattice and the elements of the array. For instance, the characteristic three-fold rotation axes of triangular arrays of nanoelements are incompatible with the two-fold symmetry of the dipolar excitations among the gaps. Note that additional frustration may be introduced in the system when the nanoelements themselves have three-fold symmetry as well. Furthermore, since the strength of the interactions among neighboring gaps depends on their areal density and inter-distances, the effect of the geometric frustration on the optical response of the array can also be modulated by changing both the size of the nanoelements and the pitch of the array.

[1]Universitat de Barcelona, Departament de Física de la Matèria Condensada, Barcelona, 08028, Spain. [2]Universitat de Barcelona, Institut de Nanociència i Nanotecnologia (IN2UB), Barcelona, 08028, Spain. [3]Institut de Ciència de Materials de Barcelona (ICMAB, CSIC), Bellaterra, 08193, Spain. [4]Institut de Microelectrònica de Barcelona (IMB-CNM, CSIC), Bellaterra, 08193, Spain. Correspondence and requests for materials should be addressed to A.L. (email: amilcar.labarta@ub.edu)

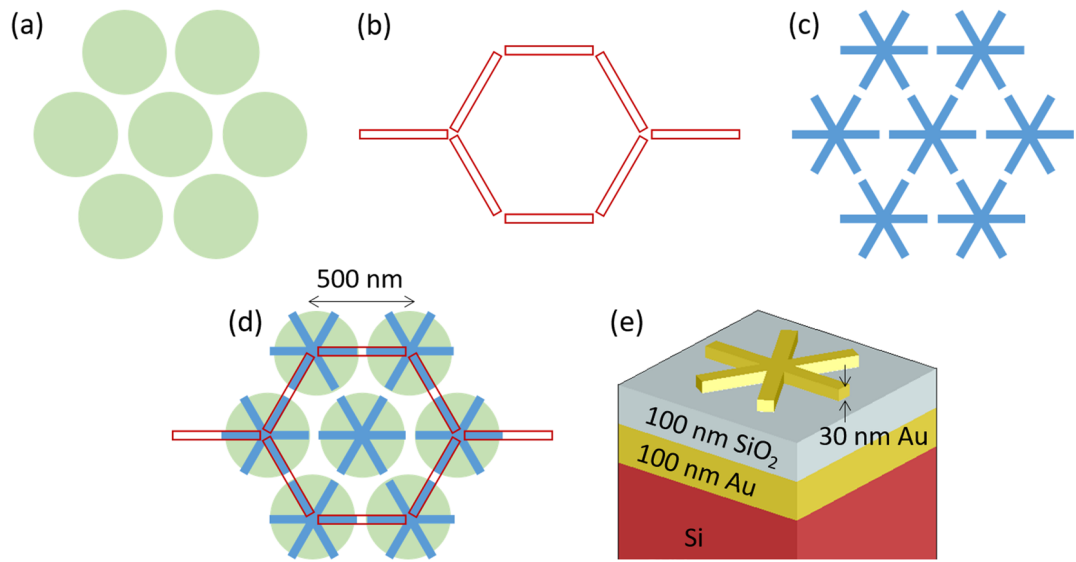

**Figure 1.** Hexagonal arrays. (**a**) Hexagonal array of disks of 450 nm in diameter. (**b**) Honeycomb array of bars of 450 nm × 45 nm. (**c**) Hexagonal array of asterisks, formed by crossing bars of 450 nm × 45 nm. (**d**) Superposition of the three arrays showing their comparable dimensions. (**e**) Substrate design for all the lattices: Si substrate with a 100 nm thick Au layer and a 100 nm $SiO_2$, on top of which 30 nm thick Au nanoelements are placed.

Studies on the geometric dependence of the plasmonic response of arrays of nanoparticles and nanostructures have already been reported by various groups[6–10]. Initially, most of the studies exploited the fundamental coupling through the gaps between close-lying simple structures. However, few works studied the dependence of the system's response on the lattice symmetry, and in particular, the excitation of higher order modes due to the interplay between low energy and high energy modes within those lattices[8]. For instance, Humphrey and Barnes studied Ag disks on glass substrates in several lattices with periods on the order of the localized surface resonances (LSR) of the particles, finding surface lattice resonances (SLR) for all types of lattices, even though the diffraction edge was the same for all of them[6]. Guo *et al.* analyzed the response of Ag nanoparticles as a function of the incident angle of the excitation, finding remarkably different extinction dispersion, which were dependent on the polarization of the incoming wave[7].

Besides, lately, perfect plasmonic absorbers have been a focus of attention due to their possible applications in photovoltaics, sensing, etc[11–13]. The use of a metal-insulator-metal (MIM) configuration was proved to be useful to obtain broadband response, localized absorption or non-angular dependence. For example, Chen *et al.*[8] investigated the response of Au nanodisks in MIM configuration with a thin insulator layer (28 nm in thickness), in such a way that Localized Surface Plasmons (LSP) and Surface Plasmon Polaritons (SPP) could couple. They found that, while resonances due to the LSP did not change significantly, the SPP modes were certainly affected by the lattice, being the honeycomb lattice the one with the richest absorption characteristics[8]. It is worth noting that the aforementioned studies focus only on the study of SLR. However, here, we go one step further and exploit multiple coupling to obtain an extended time response[14]. Although lifetimes of about 50 fs have already been reported in other lattices[15], our approach has the advantage that the relaxation of the optical excitations occurs through the echoed excitation of SLR modes at significantly longer times than in lattices without frustration. In particular, in this paper, we compare the plasmonic properties of various hexagonal arrays with pitch sizes on the order of the plasmonic resonance frequency. The lattices are formed by Au nanoelements, such as bars, disks, and asterisks, where SLR modes are remarkable over the whole time response of the system. We show that this behavior is also shown by other arrangements without triangular symmetry such as a square array of bars.

## Design

Figure 1a–c shows the design of a hexagonal array of Au disks (Fig. 1a), a honeycomb array of Au bars (Fig. 1b), and a hexagonal array of Au asterisks formed by three crossing bars (Fig. 1c). Our structures are arranged in such a way that the elements strongly interact via near-field coupling with their nearest neighbors thanks to the small gap between them. Instead of optimizing each case, the arrays have been chosen to make their dimensions comparable (see Fig. 1d). Also, we studied structures with MIM configuration where the spacer was set to a quarter-wavelength thickness to achieve maximum absorption around the plasmon resonance for normal incidence thanks to interference, as expected from the classical electromagnetic wave theory[16]. Figure 1e shows the stack over which the array is placed: Si is used as substrate, on top of which 100 nm-thick layers of both Au and $SiO_2$ are consecutively placed. The out-of-plane thickness of the structures is 30 nm in all three cases.

## Simulation Results and Discussion

Finite-Difference Time Domain (FDTD) simulations were performed with the Lumerical FDTD Solutions package[17] to get the absorption, the electric field and charge distributions, and the time evolution of the system.

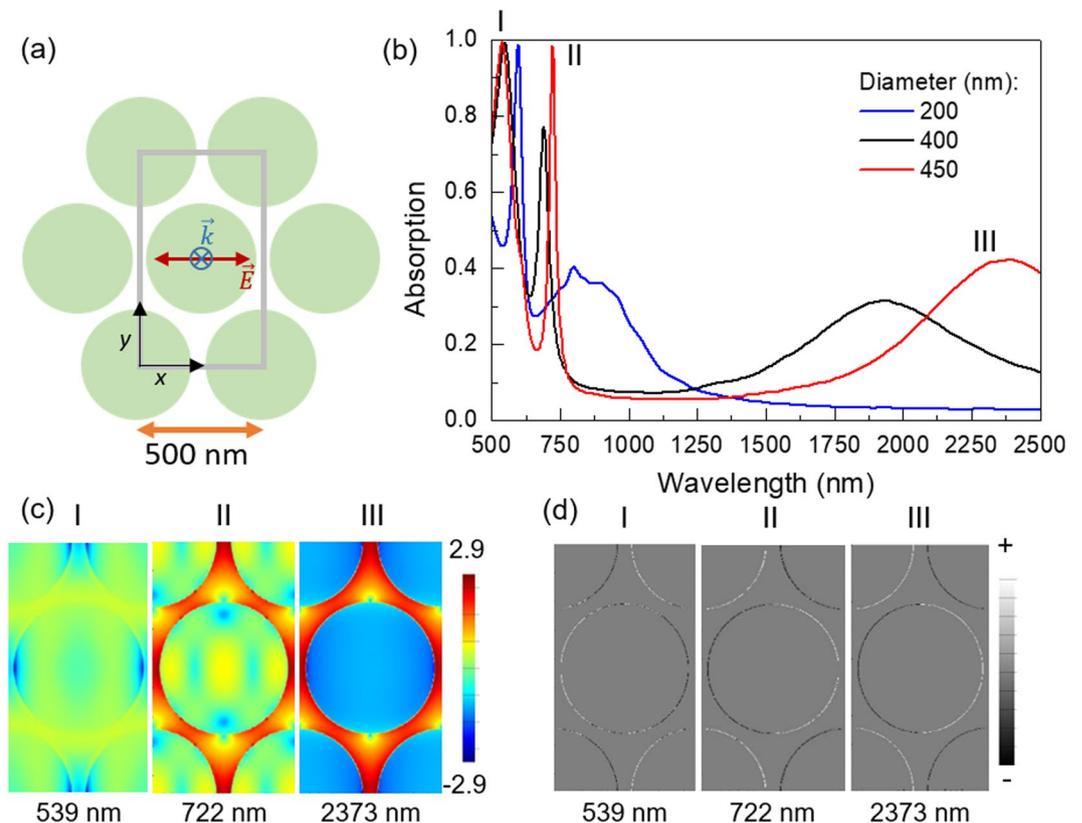

**Figure 2.** Hexagonal arrays. (**a**) Scheme of the hexagonal lattice of disks, where the grey line shows the simulation area, the black lines show the *x*- and *y*-axes and the polarization of the incoming wave has been depicted. (**b**) FDTD simulated absorption spectra of three hexagonal lattices of disks with diameters of 200 (blue), 400 nm (black), and 450 nm (red) all three with the same pitch and thickness of 500 nm and 30 nm, respectively. (**c**) $\log(|E|^2/|E_0|^2)$ and (**d**) charge distribution for the peaks in (**b**) for the 450 nm diameter case, taken at $z = 15$ nm, just at half of the thickness of the nanoelements, being the origin ($z = 0$) located at the top of the dielectric layer.

**Hexagonal array of disks.** To explore the effects of the geometrical frustration on the optical response of an ordered array of plasmonic nanoelements, a hexagonal lattice of disks placed at the vertices of the equilateral triangles forming the array (see Fig. 2a) was studied. This lattice can be also viewed as two shifted honeycomb lattices. Figure 2b shows the absorption spectra for three cases corresponding to disk diameters of 200 (blue), 400 nm (black) and 450 nm (red), all of them with the same thickness (30 nm) and a fixed 500 nm distance between the centers of neighboring disks. In this way, we expect to cover all possible cases, from structures that are strongly coupled through near-field interactions to systems with weakly interacting neighboring elements.

All the spectra show analogue features: a low-energy broad peak and two sharper high-energy peaks in the visible region of the electromagnetic (EM) spectrum that tend to superimpose on each other as the gap between the disks increases. The broad peak in the near infrared (NIR) is strongly dependent on the diameter of the structures, shifting from ca. 2373 to ca. 1931 and 874 nm when the diameter of the disks is decreased from 450 to 400 and 200 nm, respectively, and the gaps between neighboring disks are accordingly shrunk.

The spatial distributions of both the electric field in logarithmic scale, $\log(|E|^2/|E_0|^2)$, where $|E|$ is the electric field and $|E_0|$ is the incoming field, and the charge corresponding to the wavelengths of the peaks for the 450 nm diameter disks in Fig. 2b, are shown in Figs 2c and d, respectively. The electric field of the NIR modes is concentrated in the horizontal gaps in between disks and is associated with the dipolar excitation of those gaps, whose corresponding peak is broadened by the effect of damping. This is consistent with the strong shift of the NIR peak observed as a function of the disk diameter and, hence, of the gap between neighboring structures. On the contrary, for the highest energy mode, the electric field in the horizontal gaps is almost completely suppressed and the maxima are located in the tilted gaps (at an angle with respect to the electric field of the excitation pulse). The corresponding sharper and higher energy peaks of the absorption spectrum are much less dependent on the disk diameter and are presumably SLR modes enhanced because of the cavity formed by the MIM stack. The constructive interference gives rise to high absorption peaks (99.4% at 539 nm, 99.3% at 548 nm and 98.7% at 598 nm for the 450, 400 and 200 nm diameter systems, respectively). It is worth noting that the 450 nm-diameter case shows almost perfect absorption (98.4%) also at 722 nm.

Geometrically frustrated magnetic systems are known to show fluctuations over a large number of quasi-degenerate states, resulting in slow dynamics of the magnetization[18–20]. Hence, the time evolution of the optical response of the system is also analyzed to see whether frustration of dipolar excitations of the gaps between neighboring elements may give rise to an analogue behavior.

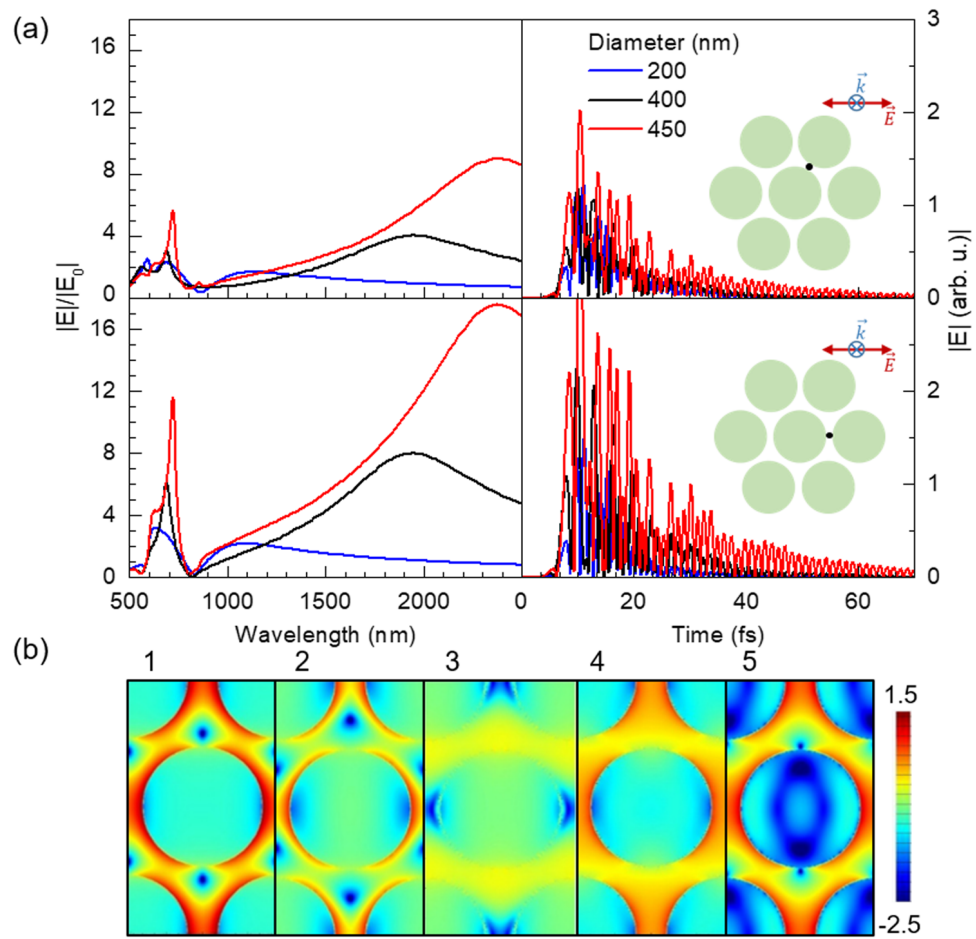

**Figure 3.** (**a**) Spectral distribution of the magnitude of the electric field, normalized to the excitation field, $|E|^2/|E_0|^2$, as a function of the wavelength (top left), and time evolution of the magnitude of the electric field (top right) both calculated at the black points depicted in the insets. Data correspond to arrays of Au disks of 200 (blue), 400 nm (black) and 450 nm (red) in diameter, with a center-to-center distance of 500 nm and a thickness of 30 nm, for an incident wave perpendicular to the $xy$ plane with the electric field parallel to the $x$-axis. (**b**) Snapshots of the time evolution of $\log(|E|^2/|E_0|^2)$ for the 400 nm case recorded at a height of $z=15$ nm with respect to the $xy$ plane, being the origin ($z=0$) located at the top of the dielectric layer. The times corresponding to snapshots 1–5 are 10.0, 10.5, 11.4, 12.5 and 13.0 fs, respectively.

Figure 3a shows the magnitude of the electric field over time at two characteristic points of the lattice depicted in the insets. Under the excitation of the system by a light pulse of only about 8 fs long, the system remains active for over 70 fs. Furthermore, the time evolution at the two feature points in Fig. 3a follows an oscillating slow decaying behavior very similar for both cases, but with higher amplitude in the horizontal gaps (parallel to the electric field of the excitation pulse), as one could expect.

To gain further insight into the time evolution of the system, Fig. 3b shows some characteristic snapshots of a sequential time evolution of $\log(|E|^2/|E_0|^2)$. The excitation source is an incident pulse along the perpendicular direction to the $xy$ plane of the array and with the electric field parallel to the $x$-axis. As a result, the system fluctuates between near-field configurations (panels 1 and 5), where the electric field is enhanced within the gaps along the $x$-axis, and surface lattice modes (panel 3) where the dipolar excitation of the horizontal gaps is suppressed. The amplitude of the oscillations over time of the magnitude of the electric field largely increases as the gaps between neighboring disks become narrower, and at the same time, the high energy peak in the spectral distributions sharpens and grows higher. This is a striking consequence of the strengthening of the near-field interactions throughout the system.

**Honeycomb array of bars.** Another realization of a plasmonic arrangement with triangular symmetry is a honeycomb array consisting of elongated bars placed along the edges of the hexagons at the midpoints of a honeycomb lattice (see Fig. 4a). This lattice may be, in fact, regarded as two interpenetrated hexagonal lattices. It is worth noting that this lattice has three bars converging in each node, which is not compatible with a simple dipolar polarization of the bars in a similar manner to the case of a spin ice[3,4]. In comparison to the previous case, one could profit from the lightning rod effect[21] and give rise to a larger enhancement of the electric field. Concomitantly, the gaps between neighboring structures are better defined as compared to the hexagonal lattice of disks and the near-field coupling between them is more directional and confined.

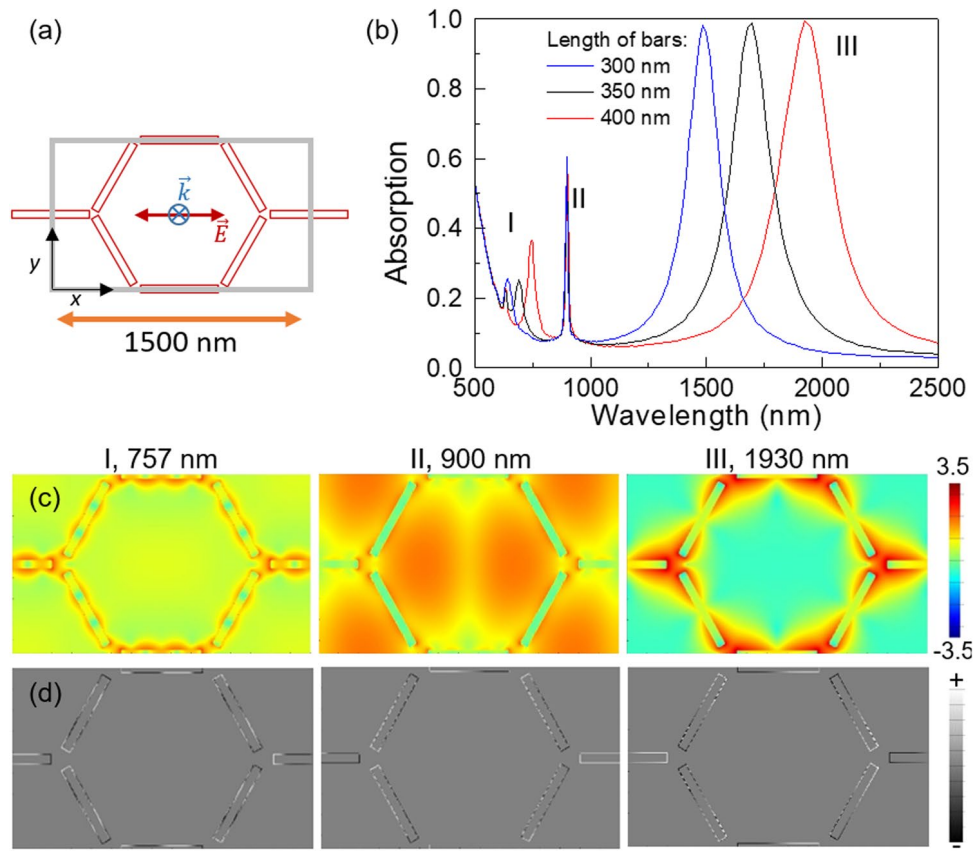

**Figure 4.** (**a**) Scheme of a honeycomb lattice of bars. The grey line shows the simulation region, the black lines show the *x*- and *y*-axes and the polarization of the incoming wave has been depicted. (**b**) Absorption spectra for lattices of 1500 nm pitch, 45 nm width and 30 nm thickness and different lengths: 300 nm (blue), 350 nm (black) and 400 nm (red). (**c**) $\log(|E|^2/|E_0|^2)$ and (**d**) charge distribution of the 400 nm bars lattice for the peaks labeled as I–III in (**b**) at $z = 15$ nm, being the origin ($z = 0$) located at the top of the dielectric layer.

Figure 4a shows the scheme of a honeycomb lattice of bars. Figure 4b shows the absorption spectra for three particular cases with 1500 nm/45 nm/30 nm (pitch/width/thickness) and different lengths: 300 nm (blue), 350 nm (black) and 400 nm (red). Unlike the previous case, this system presents a quite sharp peak in the NIR corresponding to a dipolar excitation of the bars, with almost perfect absorption (see Fig. 4b, III). This peak corresponds to a mode where all the bars show dipolar polarization giving rise to an electric field distribution that is essentially confined in the gaps between bars with a greater intensity than in the case of the disks (see Fig. 4c,d, III). This is consistent with the blue shift observed in the NIR peak for a decreasing length of the bars (see Fig. 4b). The geometric frustration of the hexagonal lattice for the dipolar excitation of the gaps is clearly evident from the charge distribution of the three bars converging on each vertex when this low-energy mode is excited: the two tilted bars show a charge accumulation at their ends with opposite sign to that at the end of the horizontal one for every junction (see Fig. 4d, III). Remarkably, this behavior strongly resembles the magnetic charge arrangements observed in magnetic spin-ice lattices[22,23]. The peaks labeled as I in Fig. 4b are of a similar nature than that of the former but with a higher order of polarization of the bars and hence they are also blue shifted as the gaps are increased. It is worth noting that the excitation of this high-energy mode is probably made possible by retardation effects associated with the relatively long length of the bars as compared to the wavelength of the electromagnetic wave.

This is consistent with our assumption that having bars instead of disks may result in a directional coupling and higher electric field enhancements. Nevertheless, the response in the visible is less intense than that of the disks, showing sharp peaks without any detectable dependence on the bar length but poor absorption. This sharper peak corresponds to a dipolar excitation along the horizontal bars and a kind of dipolar excitation of the tilted bars in the transverse direction, but modified in the ends due to the constraint of the 3-bars junction. This leads to accumulation of charges of the same sign in the two tilted bars and of opposite sign in the horizontal one (see Fig. 4d, II). As a result, the electric field map shows big lobes of electric field enhancement surrounding the tilted bars (see Fig. 4c, II). These modes in the visible, showing an enhancement of the electric field far beyond the gaps, are assigned to SLR.

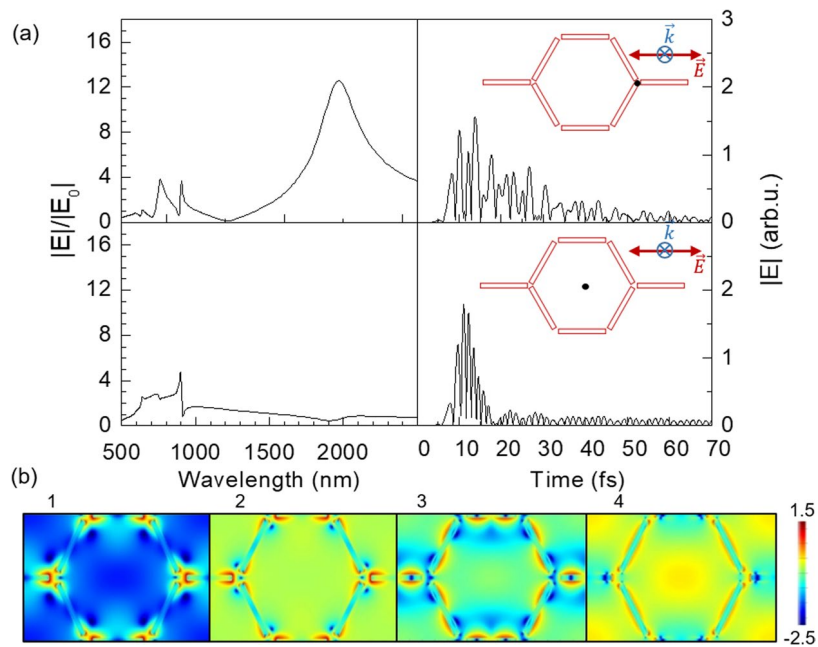

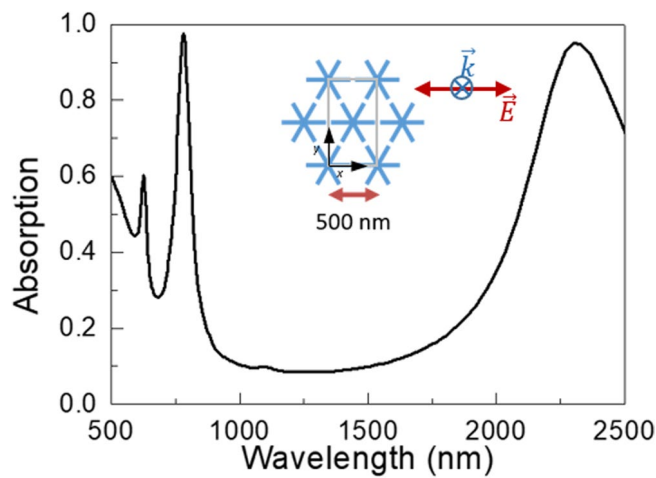

**Figure 5.** (**a**) Electric field as a function of wavelength (spectral distribution) (left) and time (right), for the characteristic black points shown in the insets. (**b**) Frames corresponding to the time evolution of $\log(|E|^2/|E_0|^2)$ at $z = 15$ nm, being the origin ($z = 0$) located at the top of the dielectric layer. The snapshots 1–4 correspond to times 8.1, 8.3, 9.1 and 10.0 fs, respectively.

**Figure 6.** FDTD simulated absorption spectra for a hexagonal lattice of asterisks with dimensions of 500 nm/450 nm/45 nm/30 nm (pitch/length/width/thickness). The inset shows the scheme of the lattice, the black lines show the $x$- and $y$-axes and the polarization of the incoming wave has been depicted.

Analyzing the time response of the system (see Fig. 5a), we observe that the evolution of the field intensity in the gaps is pretty similar to that of the disks, with a slow decay of the intensity that extends over 70 fs. Analyzing the time-lapse frames (see Fig. 5b), similarly to the hexagonal lattice of disks, the state of the system changes in a quasi-periodic manner between LSR where the electric field is concentrated in the gaps and SLR where the electric field is enhanced around the long edges of the bars and almost suppressed in the gaps. Moreover, it is remarkable that, at certain times, a noticeable enhancement of the electric field takes place in the center of the hexagons. Interestingly, when the detector is placed at the very centre of the lattice (Fig. 5a, bottom panels) even though the enhancement of the electric field is actually low, the main contributions to the spectral distribution are in the visible range and correspond to the excitations of SLR that are still shining after relatively long times.

These results are in good agreement with the previously reported response of a hexagonal lattice of asterisks[5], which also shows a broadband in the NIR attributed to the dipolar excitation of the gaps between neighboring asterisks, and higher order modes, lying on the visible, associated with collective modes induced by the ordered array (see Fig. 6).

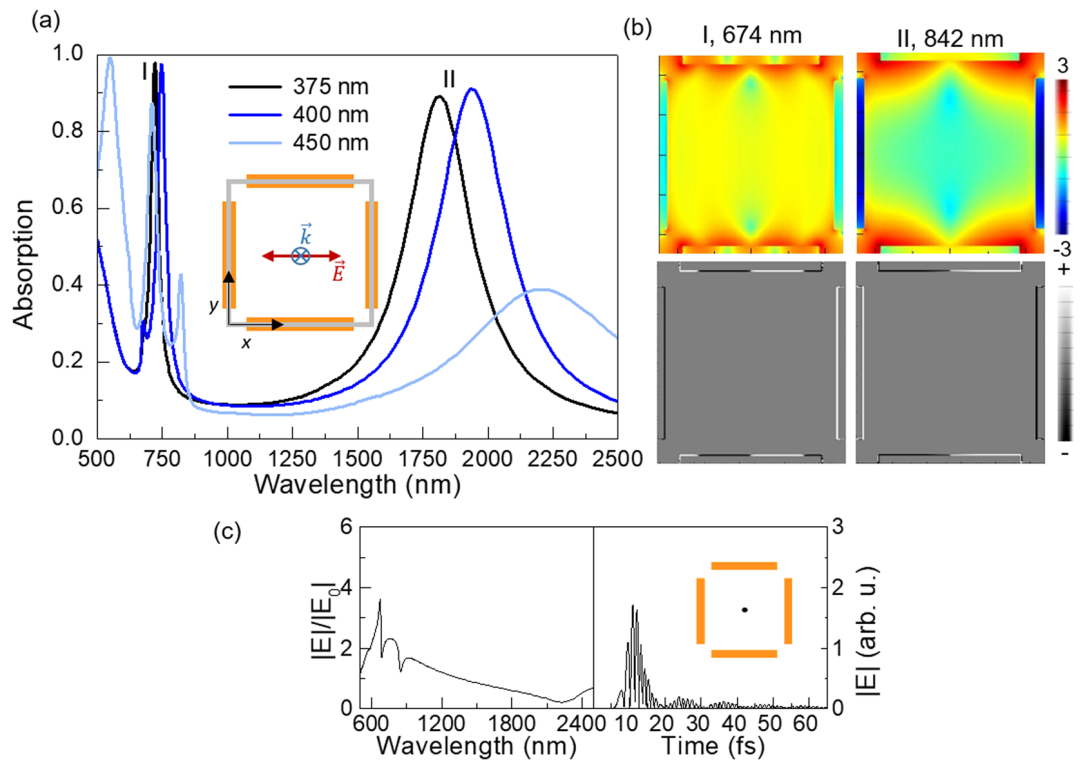

**Figure 7.** (**a**) FDTD simulated absorption spectra of several square lattices of bars (scheme of the lattice as inset), of 500 m/45 nm/ 30 nm (pitch/width/thickness) and various lengths: 375 nm (black), 400 nm (dark blue) and 450 nm (light blue). (**b**) $\log(|E|^2/|E_0|^2)$ (top colored panels) and charge distribution (bottom grey-scale panels) of the 375 nm bars lattice for the peaks labeled as I-II in a at $z = 15$ nm, being the origin ($z = 0$) located at the top of the dielectric layer. (**c**) Electric field as a function of wavelength (spectral distribution) (left) and time (right) in the center of the unit cell, depicted with a black point in the inset.

All in all, we have a system which presents almost perfect absorption in the NIR region and high electric field enhancements over a much longer time than the duration of the incident pulse, even though the response in the visible range is moderate. However, the latter may be improved optimizing the geometric parameters of the elements and the thickness of the dielectric spacer.

### Other lattices with geometric frustration

Geometric frustration for the dipolar excitation is not exclusive of triangular lattices since simulations of other lattices show also a similar interplay between SLR and LSR modes. Here, we present the case of a square array of bars (see inset in Fig. 7a). Figure 7a shows the absorption spectra for arrays of 500 nm/45 nm/30 nm (pitch/width/thickness) and lengths ranging from 375 to 450 nm. For all lengths, the array presents a broad peak in the NIR region of the spectrum and sharper peaks in the visible, as seen as well in the honeycomb lattice. This is because full ferroelectric polarization of the array is geometrically hindered by the perpendicular orientation of the bars of the two interpenetrated square lattices that form the array. However, it is interesting to notice the reduced absorption in the visible for the 450 nm case when the gap between perpendicular and horizontal bars is small, leading to higher frustration. When the gaps widen, near-field coupling in the gaps decreases and the absorption in the NIR increases.

Figure 7b shows the electric field ($\log(|E|^2/|E_0|^2)$) (top panels) and charge (bottom panels) spatial distributions for the 375 nm case, corresponding to the maximum absorption, indexed as I-II in Fig. 7a, showing the dipolar polarization of the horizontal bars for the lowest energy mode as compared to the higher complexity excitation of the system for the peak in the visible.

Moreover, the time response of this lattice, displayed in Fig. 7c, shows also a similar behavior to the one observed in the triangular lattices. This is due to the orthogonal arrangement of nanoelements at every vertex of the lattice. The coexistence of those junctions of two well-oriented bars (parallel to the linearly polarized excitation source) and those of two bars perpendicular to the latter (hence more difficult to be excited), leads as well to an interplay between localized modes with dipolar excitation of the gaps and collective modes of the array.

Consequently, geometric frustration shows up in the optical response of a plasmonic array when the gaps among neighboring nanoelements are not aligned along a unique spatial axis. In this case, the dipolar coupling between neighboring structures cannot be simultaneously fulfilled for all of them. This is obviously the case not only for triangular arrays, but also for many other arrays, such as for example a square lattice of rectangular bars. However, we would like to emphasize that the enhancement of the excitation of collective modes in triangular arrays can be significantly larger because of the inherent geometric incompatibility of dipolar excitations and three-fold symmetry axes.

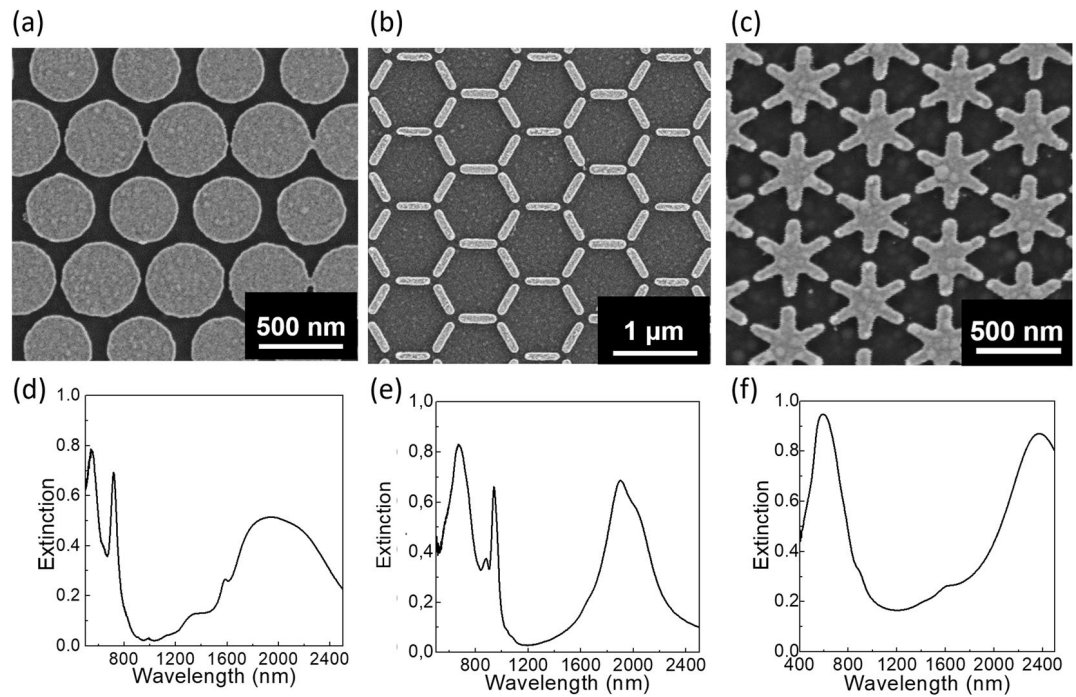

**Figure 8.** (**a**–**c**) SEM micrographs and (**d**–**f**) FTIR extinction (1-$R$-$T$ and with $T=0$) spectra under unpolarized light, for a hexagonal lattice of disks with a pitch of 500 nm (**a**,**d**), a honeycomb lattice of bars with a pitch of 1500 nm (**b**,**e**), and a hexagonal lattice of asterisks with a pitch of 500 nm (**c**,**f**), all of them fabricated in MIM configuration with a $SiO_2$ thickness of 100 nm and metallized using 0.5 nm of Cr and 20 nm of Au.

### Experimental Results and Discussion

A set of samples of triangular symmetry was fabricated in MIM configuration with a 100 nm-thick $SiO_2$ spacer layer and 20 nm-thick nanostructures. For all samples, a 0.5 nm Cr layer was deposited as adhesion promoter of the elements to the substrate. Electron beam lithography (EBL) was chosen to get fast prototyping to study effects of different geometries in the plasmonic response.

Figure 8 shows one example of the fabricated arrays for each lattice type and their corresponding extinction (1-$R$-$T$, $R$ being the reflectance and $T$ the transmitance and with $T=0$) spectra measured using a FTIR (Fourier Transformed Infrared Spectroscopy) spectrophotometer attached to an optical microscope in reflection mode. The optical response of the asterisks can be divided in two regimes: in the visible part, intense narrow peaks can be distinguished and in the NIR part, a broad peak is observed. The experimental spectra (Fig. 8d–f) are in qualitative agreement with the simulated spectra (Figs 2b, 4b and 6). The modulation of the broad peak in the NIR can be attributed to experimental imperfections of the structure like small variations in size and some other fabrication defects such as the occurrence of elements connected to some of their neighbors, rounded shapes, non-uniform thickness of the elements, etc. For the honeycomb lattice, the main difference with respect to the simulations is the higher intensity of the peaks in the visible region that may be due to the different element size and hence different MIM cavity response than for the simulated case. Finally, for the asterisk lattice the two peaks in the visible are collapsed into one, which seems to be a consequence of the adhesion layer, together with the natural broadening of the peaks as compared to simulations due to defects such as roughness. However, despite all these differences, the measurements reveal a good agreement with the FDTD simulations.

### Conclusions

In summary, we have explored the optical response of four types of arrays of plasmonic nanoelements, three with triangular symmetry and one of square symmetry, all of them presenting features that are a consequence of the geometric frustration for the dipolar excitation through near-field interactions of the gaps among neighboring nanoelements.

The geometries studied herein exhibit perfect absorption peaks, which are tunable with the design parameters. While the hexagonal lattice of disks shows two high absorption peaks in the visible and the honeycomb lattice only displays high absorption in the NIR, the array of asterisks and the square lattice of bars achieve almost perfect absorption both in the visible and NIR regions. Bars and asterisks are easier to polarize giving rise to a higher confinement of the electric field within the gaps as compared to disks, which have to be very large in order to achieve near-field coupling at the expense of almost covering the whole dielectric surface.

All lattices show extended time response with echoed excitation of the collective modes that remains excited at long times due to the frustration of the dipolar polarization of the gaps. Consequently, they are suitable for the maximization of both the absorption and ulterior radiation of the impinging electromagnetic wave, resulting in higher energy radiation at shorter wavelengths.

In addition, the enhancement of the electric field associated with the optical response takes place over most of the area of the array as the time elapses, not being limited to hot spots as is the case in dimer antennas, thus avoiding the issue of very precise positioning of the target molecules[14]. As a result, although the maximum enhancement of the electric field is smaller than in the case of dimer antennas, frustrated plasmonic arrays may allow increasing the detection limit by interacting with larger number of biomolecules.

All in all, the delocalized nature of the collective modes, lying in the visible range of the spectrum, together with the extended time response make these systems suitable for enhanced spectroscopies, such as Raman[24] or IR[25], or photovoltaic applications[26].

## Methods

**Simulations.** Finite Difference Time Domain (FDTD) simulations were done using the FDTD Solutions Package from Lumerical[17]. For all cases, the simulation area comprised one unit cell. A short-time pulse with linear polarization parallel to the $x$-axis was injected perpendicular to the substrate ($z$-axis) in order to excite the system. Perfectly matched layers were set in the $z$-axis while for the $x$- and $y$- axis, periodic boundary conditions were imposed. In the simulations, data from Johnson and Christy[27] were used for the permittivity of Au, whereas for $SiO_2$ and Si, the data from Palik[28] were adopted.

**Sample fabrication.** Nanometer scale structures forming arrays with triangular symmetry were fabricated in samples having a MIM configuration using a Si wafer metallized with 5 nm of Ti as adhesion promoter and a 60 nm thick Au layer, acting as mirror, on top of which a 100 nm thick $SiO_2$ layer was deposited by Plasma Enhanced Chemical Vapor Deposition (PECVD). Poly (methyl methacrylate) (950 PMMA A2, MicroChem) was spin-coated at 1200 rpm for 1 minute onto the substrate and cured at 180 °C for 1 minute. EBL was done with a 20 μm gun aperture, a step size of 10 nm and a 20 kV voltage, using the Raith Two 150. The dose was fixed to 180 μC/cm² and the dose factor was changed from area to area to finely tune the size. After the exposure, the sample was developed by dipping it in a 1:3 solution of methyl isobutyl ketone (MIBK) in isopropanol (IPA) for 30 seconds and then rinsed in IPA for 30 seconds to stop the reaction. Electron beam evaporation (ATC Orion, AJA International, Inc) was used to metallize the sample. Finally, lift-off was performed in an acetone bath with ultrasounds at 40 °C.

**Optical characterization.** Extinction spectra (1-$R$) were calculated from the measured reflectance, $R$, and considering zero transmittance. The reflectance from the films was calculated using an FTIR spectrophotometer (Vertex 70, Bruker) attached to an optical microscope (Hyperion) and a 4X objective and a spatial mask. For this purpose, unpolarized light was used.

## Data Availability

The data generated during this study are available from the corresponding author.

### Acknowledgements

This work was supported by Spanish MINECO (MAT2015-68772- P, BES-2013-065377), Spanish ICTS Network MICRONANOFABS, partially supported by MINECO, and European Union FEDER funds. A.M. acknowledges funding from the European Research Council (ERC) grant agreement 637116.

### Author Contributions

A.C.R. did the simulations and fabricated the samples. A.C.R., A.F.R., X.B. and A.L. discussed the simulation results. F.P.M. supervised the fabrication process. A.C.R., A.E. and A.M. performed the optical characterization. A.C.R. and A.L. wrote the manuscript with input from all authors. All authors have given approval to the final version of the manuscript.

### Additional Information

**Competing Interests:** The authors declare no competing interests.

**Publisher's note:** Springer Nature remains neutral with regard to jurisdictional claims in published maps and institutional affiliations.